\documentclass[runningheads]{llncs}

\usepackage{isabelle,isabellesym}
\usepackage{amssymb}
\usepackage{xcolor}
\definecolor{keyword}{RGB}{215, 58, 73}
\definecolor{command}{RGB}{215, 58, 73}
\definecolor{math}{RGB}{0, 92, 197}
\newcommand{\mathstart}{\color{math}\it}
\newcommand{\mathend}{\normalcolor\it}
\let\oldisabellestyleit\isabellestyleit
\renewcommand{\isabellestyleit}{%
\oldisabellestyleit%
\def\isachardoublequoteopen{{\color{math}\scriptsize\it\texttt{"}}\aftergroup\mathstart}%
\def\isachardoublequoteclose{{\color{math}\scriptsize\it\texttt{"}}\aftergroup\mathend}%
\def\isacharbackquote{\rm\`{}}
\def\isacharat{\raisebox{-0.025em}{\textit{\scriptsize@}}}%
\def\isacharunderscore{\texttt{\_}}%
\def\isacharunderscorekeyword{\texttt{\_}}%
}
\isabellestyle{it}
\renewcommand{\isastyle}{\it\color{black}}
\renewcommand{\isastyletext}{\rm\color{black}}
\renewcommand{\isastyletxt}{\rm\color{black}}
\renewcommand{\isastylecmt}{\rm\color{black}}
\let\oldisakeyword\isakeyword
\renewcommand{\isakeyword}[1]{{\color{keyword}\oldisakeyword{#1}}}
\renewcommand{\isacommand}[1]{{\color{command}\oldisakeyword{#1}}}

\usepackage{graphicx}
\usepackage{hyperref}
\usepackage{breakurl}

\begin{document}
\title{Formal specification of a security framework for smart contracts}
%
%
\author{Mikhail Mandrykin\inst{1} \and
Jake O'Shannessy \inst{2} \and
Jacob Payne\inst{2} \and
Ilya Shchepetkov\inst{1}}
\authorrunning{Mandrykin et al.}
%
\institute{
ISP RAS, Moscow, Russia\\
\email{\{mandrykin,shchepetkov\}@ispras.ru}
\and
Daohub, San Francisco, USA\\
\email{joshannessy@gmail.com, jacob@daohub.io}}
\maketitle              
\begin{abstract}
As smart contracts are growing in size and complexity, it becomes harder and harder to ensure their correctness and security.
Due to the lack of isolation mechanisms a single mistake or vulnerability in the code can bring the whole system down, and due to this smart contract upgrades can be especially dangerous.
Traditional ways to ensure the security of a smart contract, including DSLs, auditing and static analysis, are used before the code is deployed to the blockchain, and thus offer no protection after the deployment.
After each upgrade the whole code need to be verified again, which is a difficult and time-consuming process that is prone to errors.
To address these issues a security protocol and framework for smart contracts called Cap9 was developed.
It provides developers the ability to perform upgrades in a secure and robust manner, and improves isolation and transparency through the use of a low level capability-based security model.
We have used Isabelle/HOL to develop a formal specification of the Cap9 framework and prove its consistency.
The paper presents a refinement-based approach that we used to create the specification, as well as discussion of some encountered difficulties during this process.

\end{abstract}
\section{Introduction}
Ethereum~\cite{Ethereum_white_paper} is a global blockchain platform for decentralised applications with a built-in Turing-complete programming language.
This language is used to create smart contracts~--- automated general-purpose programs that have access to the state of the blockchain, can store persistent data and exchange transactions with other contracts and users.
Such contracts have a number of use-cases in different areas: finance, insurance, intellectual property, internet of things, voting, and others.

However, creating a \textit{reliable} and \textit{secure} smart contract can be extremely challenging. Ethereum guarantees that the code of a smart contract would be executed precisely as it is written through the use of a consensus protocol, which resolves potential conflicts between the nodes in the blockchain network.
It prevents malicious nodes from disrupting and changing the execution process, but does not protect from the flaws and mistakes in the code itself.
And due to the lack of any other control on the execution of the code any uncaught mistake can potentially compromise not only the contract itself, but also other contracts that are interacting with it and expect a certain behavior from it.

Such flaws can be turned into vulnerabilities and cause a great harm, and there are many examples of such vulnerabilities and attacks that exploit them~\cite{atzei_survey_2017}.
Developers can ensure the security of a contract using auditing, various static analysis tools~\cite{luu_making_2016,kalra_zeus:_2018}, domain-specific languages~\cite{ergo_dsl,frantz_institutions_2016}, or formal verification~\cite{bhargavan_formal_2016}.
These are excellent tools and methods that can significantly improve the quality of the code.
But they are not so effective during the upgrades, which is a common process for almost every sufficiently sophisticated smart contract.
Upgrades are necessary because it is the only way to fix a bug that was missed during the verification process.
However, they can also introduce their own bugs, so after each upgrade the code needs to be verified again, which may cost a lot of time and effort.

These issues are addressed by the Cap9 framework~\cite{cap9_white_paper}.
It provides means to isolate contracts from each other and restrict them from doing dangerous state-changing actions unsupervised, thus greatly reducing risks of upgrades and consequences of uncaught mistakes.
Cap9 achieves this by using a low level capability-based security model, which allows to explicitly define what can or can not be done by any particular contract.
Once defined, such capabilities, or permissions, are visible to anyone and can be easily understood and independently checked, thus increasing transparency of the system.

In order to be trusted, the Cap9 framework itself needs to be formally verified.
The specification of the framework must be formalised and proved, in order to show that it is consistent and satisfies the stated properties.
Then the implementation, which is a smart contract itself, must be proved to be compliant with its specification.
In this paper we are focusing only on the first part~--- on developing and proving a formal specification of the Cap9 framework using the Isabelle/HOL theorem prover~\cite{nipkow_isabelle_2002}
The paper presents a refinement-based approach that we used to create the specification, and evaluates the chosen formal method by describing encountered difficulties during this process.

The following section outlines the features and capabilities of the Cap9 framework.
Section 3 presents the Isabelle/HOL specification, as well as the difficulties we have encountered and the refinement process we used to develop it.
Related work is reviewed in Section 4.
The last section concludes the paper and considers future work.

\section{Cap9 Framework}
The Cap9 framework achieves isolation by interposing itself between the smart contracts that are running on top of it and potentially dangerous actions that they can perform, including calling other smart contracts, writing to the storage and creating new contracts.
Such actions can be performed only using special ``System Call'' interface provided by the framework.
Via this interface it has complete control over what contracts can and cannot do.
Each time a system call is executed Cap9 conducts various runtime security checks to ensure that a calling contract indeed has necessary rights to perform a requested action.
It works similar to how operating system kernels manage accesses of programs to the hardware (like memory and CPU) and protect programs from unauthorised accesses to each other.


In order to ensure that a contract correctly uses the system call interface and does not perform any actions bypassing the framework its source code needs to be verified.
Cap9 does it on-chain and it checks that the source code does not contain any forbidden instructions, like ones allowing to make state changes, make an external call, or self destruct. The check is called procedure bytecode validation. The valid code is essentially only allowed to perform local computations (those not involving any calls or modifications of the store) and delegate calls to a special predefined kernel address. This is a very simple property that can be ensured by an efficient dynamic check that is performed only once upon the registration of the newly deployed code.
Once the code is validated the corresponding contract can be registered in the framework as a \textit{procedure} and thus access its features.

There are system calls available to securely perform the following actions:
\begin{itemize}
    \item Register new procedure in the framework;
    \item Delete a registered procedure;
    \item Internally call a registered procedure;
    \item Write data to the storage;
    \item Append log record with given topics;
    \item Externally call another Ethereum address and/or send Ether;
    \item Mark a procedure as an \textit{entry} procedure~--- one that would handle all the incoming external calls to this contract system or organisation.
\end{itemize}

As a typical smart contract, Cap9 has access to the storage~--- a persistent 256 x 256 bits key-value store.
A small part of it is restricted and can be used only by the framework itself.
It has a strict format and is used to store the list of registered procedures, as well as procedure data, addresses of entry and current procedures and the Ethereum address of the deployed framework itself.
This part is called the \textit{kernel storage}.
The rest of the storage is open to use by any registered procedure either directly (in case of read) or through a dedicated system call (in case of write).

Traditional kernels have a lot of abstraction layers between programs and hardware.
Unlike them, Cap9 exposes all the underlying Ethereum mechanisms directly to the contracts, with only a thin permission layer between them.
This layer implements a capability-based access control, according to which in order to execute a system call a procedure must posses a \textit{capability} with an explicit permission.
Such capability has a strict format, which is different for each available type of system calls.

Capabilities can be used to restrict components of a smart contract system and thus to implement the principle of least privilege. They can also be used as base primitives to create a custom high-level security policy model to better fit a particular use case. Such policy would be simple to analyze and understand, but able to limit possible damage from bugs in the code or various malicious actions (including replacing the code of a contract via the upgrade mechanism).

Cap9 is compatible with both EVM and Ewasm applications.

\section{Formal Specification}
The main goal of formalizing the interface specification of the Cap9 security framework was to ensure internal consistency and completeness of its description as well as to provide a reliable reference for all of its implementations. The reference should eventually serve as an intermediate between the users and the developers of any Cap9 implementation ensuring full compatibility of all further system uses and implementations. The source specification itself is formulated as a detailed textual description of the system interface~\cite{cap9_spec}, which is language-agnostic and relies on the binary interfaces of the underlying virtual machine. Thus all the data mentioned in the specification is given an explicit concrete bit-level representation, which is intended to be shared by all system users and implementations.

\subsection{Consolidation of low-level representation with high-level semantics}
One of the immediately arising challenges of formally verifying a system with very explicit specifications on concrete data representation is efficiently establishing a correspondence between this representation and the corresponding intended semantics, which is used for actual reasoning about the system and therefore for the actual proof.

 A particular example in our case is the representation and the semantics of capability subsets. Each capability of every procedure in the system logically corresponds to a set of admissible values for some parameter configuration, such as kernel storage address (for writing to the storage), Ethereum address and amount of gas for external procedure call, log message with several special topic fields etc. Each such set is composed of a (not necessarily disjoint) union of a number of subsets, which in their turn directly correspond to some fixed representations. A subset of writable addresses, for example, is represented as a pair of the starting address and the length of a continuous range of admissible addresses. Thus the entire write capability of any kernel procedure is a union of such continuous address ranges.

 But it's important to note that while on one hand we clearly need to state the set semantics of the write capability (as a generally arbitrary set of addresses), in particular this is especially convenient semantics to be used for proofs of generic capability properties, such as transitivity; on the other hand, however, we have a clearly indicated format of the corresponding capability representation stated in the system specification, which is not a set, but a range of storage cells holding the bit-wise representations of the starting addresses and lengths of the corresponding ranges.

 If we stick with the specified representation, we will be unable to efficiently use many powerful automated reasoning tools provided with Isabelle/HOL, such as the classical reasoner and the simplifier readily pre-configured for the set operations. However, if we just use the set interpretation, the specification on the concrete representation will be notoriously hard to express. Hence we likely need several different formalizations of a notion of capability on several levels of abstraction. We actually used three representations: the concrete bit-wise representation, the more abstract representation with the length of the range expressed as natural number (and with an additional invariant), and finally the set representation. By using separate representations we ended up with small simple proofs for both generic capability properties and their concrete representations.

\subsection{Correspondence relation vs. representation function}

Eventually we decided to employ the same refinement approach with several formalizations for the entire specification, thus obtaining two representations of the whole system: the structured high-level representation with additional type invariants and the low-level representation as the mapping from 32-byte addresses to 32-byte values, i.e. the state of the kernel storage. However, using separate representations raises a problem of efficiently establishing the correspondence between them. Initially we tried a more general approach based on the correspondence relation. Yet to properly transfer properties of the high-level representation to the low-level one, the relation should enjoy at least two properties: injectivity and non-empty image of every singleton:

\smallskip
\noindent{\isacommand{lemma}\isamarkupfalse\isastyle%
\ rel{\isacharunderscore}injective{\isacharcolon}\ {\isachardoublequoteopen}{\isasymlbrakk}s\ {\isasymtturnstile}\ {\isasymsigma}\isactrlsub {\isadigit{1}}{\isacharsemicolon}\ s\ {\isasymtturnstile}\ {\isasymsigma}\isactrlsub {\isadigit{2}}{\isasymrbrakk}\ {\isasymLongrightarrow}\ {\isasymsigma}\isactrlsub {\isadigit{1}}\ {\isacharequal}\ {\isasymsigma}\isactrlsub {\isadigit{2}}{\isachardoublequoteclose}}\\
{\isacommand{lemma}\isamarkupfalse\isastyle%
\ non{\isacharunderscore}empty{\isacharunderscore}singleton{\isacharcolon}\ {\isachardoublequoteopen}{\isasymexists}\ s{\isachardot}\ s\ {\isasymtturnstile}\ {\isasymsigma}{\isachardoublequoteclose}}
\smallskip

Here $\isasymtturnstile$ stands for the correspondence relation, $\sigma$~--- for the high-level representation and $s$~--- for the concrete one. We noticed that proving the second lemma essentially requires defining a function mapping an abstract representation to the corresponding concrete one. Thus this approach results in significant redundancy in a sense that both the function defined for the sake of proving the second lemma and the correspondence relation itself repeat essentially the same constraints on the low-level representation. For a very simple example consider:

\smallskip{
\noindent\isacommand{definition}\isamarkupfalse\isastyle%
\ models\ {\isacharcolon}{\isacharcolon}\ {\isachardoublequoteopen}{\isacharparenleft}word{\isadigit{3}}{\isadigit{2}}\ {\isasymRightarrow}\ word{\isadigit{3}}{\isadigit{2}}{\isacharparenright}\ {\isasymRightarrow}\ kernel\ {\isasymRightarrow}\ bool{\isachardoublequoteclose}\ {\isacharparenleft}{\isachardoublequoteopen}{\isacharunderscore}\ {\isasymtturnstile}\ {\isacharunderscore}{\isachardoublequoteclose}{\isacharparenright}\ \isakeyword{where}\isanewline
\ \ {\isachardoublequoteopen}s\ {\isasymtturnstile}\ {\isasymsigma}\ {\isasymequiv}\ unat\ {\isacharparenleft}s\ {\isacharparenleft}addr\ Nprocs{\isacharparenright}{\isacharparenright}\ {\isacharequal}\ nprocs\ {\isasymsigma}{\isachardoublequoteclose}\\
\isacommand{definition}\isamarkupfalse%
\ {\isachardoublequoteopen}witness\ {\isasymsigma}\ a\ {\isasymequiv}\ case\ addr{\isasyminverse}\ a\ of\ Nprocs\ {\isasymRightarrow}\ of{\isacharunderscore}nat\ {\isacharparenleft}nprocs\ {\isasymsigma}{\isacharparenright}{\isachardoublequoteclose}
}\smallskip

Here not only we need to repeatedly state the relationship between the value of kernel storage at address $addr~ Nprocs$ and the number of procedures registered in the system ($nprocs~\sigma$) twice, but we also potentially have to define the address encoding and decoding functions ($addr$ and $addr\isasyminverse$) separately and to prove the lemma about their correspondence. We discuss our approach to address encoding in the following section and here only emphasize the redundancy arising from the approach based on the correspondence relation.

It also worth noting that merely transferring or lifting the properties stated for one representation to another is insufficient as we would like to also be able to conveniently represent mixed properties such as a property specifying the result of an operation on the high level, but also stating an additional constraint on its concrete representation e.g. that some irrelevant bits in the representation should be zeroed and some others remain unchanged.

At the same time, the major reason for introducing the correspondence relation instead of using a function is an inherent ambiguity of the encoding of the high-level representation into the low-level one. However, after carefully revisiting the initial specification of the system we noticed that the ambiguity of representation in our system actually arises only from the unused storage memory rather than from the presence of any truly distinct ways of representing the same state. But this particular kind of ambiguity can be efficiently expressed using a representation function with an additional parameter~--- i.~e. the state of the unused memory.

Let's illustrate our formalization approach that is based on representation functions on the example of Procedure Call capability. The specification of this capability is as follows:

\emph{The capability format for the Call Procedure system call defines a range of
procedure keys which the capability allows one to call. This is defined as a base
procedure key b and a prefix s. Given this capability, a procedure may call any
procedure where the first s bits of the key of that procedure are the same as
the first s bits of procedure key b.}

\smallskip
{\noindent\includegraphics[width=1\textwidth]{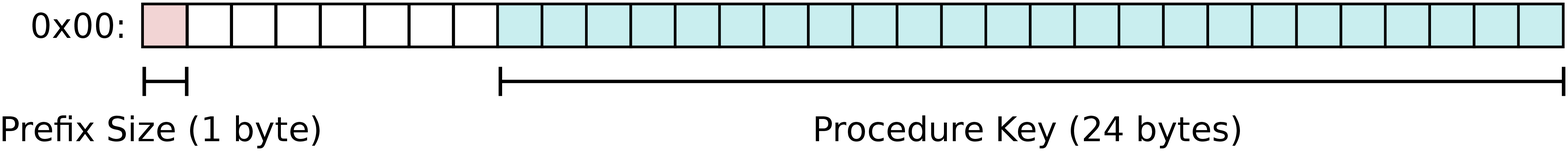}}

Here the unused space is left blank. Beforehand we strive to make the actual formulation of the arising injectivity lemma as simple as possible by eliminating premises of the lemma and turning them into type invariants. So we introduce the following definitions:\\
{\noindent\isacommand{typedef}\isamarkupfalse\isastyle%
\ prefix{\isacharunderscore}size\ {\isacharequal}\ {\isachardoublequoteopen}{\isacharbraceleft}n\ {\isacharcolon}{\isacharcolon}\ nat{\isachardot}\ n\ {\isasymle}\ LENGTH{\isacharparenleft}key{\isacharparenright}{\isacharbraceright}{\isachardoublequoteclose}\\
\isacommand{definition}\isamarkupfalse%
\ {\isachardoublequoteopen}prefix{\isacharunderscore}size{\isacharunderscore}rep\ s\ {\isasymequiv}\ of{\isacharunderscore}nat\ {\isasymlfloor}s{\isasymrfloor}\ {\isacharcolon}{\isacharcolon}\ byte{\isachardoublequoteclose}\ \isakeyword{for}\ s\ {\isacharcolon}{\isacharcolon}\ prefix{\isacharunderscore}size\\
\isacommand{type{\isacharunderscore}synonym}\isamarkupfalse\isastyle%
\ prefixed{\isacharunderscore}capability\ {\isacharequal}\ {\isachardoublequoteopen}prefix{\isacharunderscore}size\ {\isasymtimes}\ key{\isachardoublequoteclose}\\
\isacommand{definition}\isamarkupfalse~--- \emph{set interpretation of single write capability}\\
\phantom{m}{\isachardoublequoteopen}set{\isacharunderscore}of{\isacharunderscore}pref{\isacharunderscore}cap\ {\isacharparenleft}sk\ {\isacharcolon}{\isacharcolon}\ prefixed{\isacharunderscore}capability{\isacharparenright} \ {\isasymequiv}\ let\ {\isacharparenleft}s{\isacharcomma}\ k{\isacharparenright}\ {\isacharequal}\ sk\ in\\\ \phantom{m} {\isacharbraceleft}k{\isacharprime}\ {\isacharcolon}{\isacharcolon}\ key{\isachardot}\ take\ {\isasymlfloor}s{\isasymrfloor}\ {\isacharparenleft}to{\isacharunderscore}bl\ k{\isacharprime}{\isacharparenright}\ {\isacharequal}\ take\ {\isasymlfloor}s{\isasymrfloor}\ {\isacharparenleft}to{\isacharunderscore}bl\ k{\isacharparenright}{\isacharbraceright}{\isachardoublequoteclose}\\
\isacommand{adhoc{\isacharunderscore}overloading}\isamarkupfalse\isastyle%
\ rep\ prefix{\isacharunderscore}size{\isacharunderscore}rep~~--- prefix\_size\_rep \emph{is now denoted as} {\isasymlfloor}$\cdot${\isasymrfloor}\\
\isacommand{definition}\isamarkupfalse\isastyle~--- \emph{low-level (storage) representation of single write capability}\\
\phantom{i}\ {\isachardoublequoteopen}pref{\isacharunderscore}cap{\isacharunderscore}rep\ {\isacharparenleft}sk\ {\isacharcolon}{\isacharcolon}\ prefixed{\isacharunderscore}capability{\isacharparenright} r\ {\isasymequiv}\ let\ {\isacharparenleft}s{\isacharcomma}\ k{\isacharparenright}\ {\isacharequal}\ sk\ in\\\ \phantom{m}{\isasymlfloor}s{\isasymrfloor}\ \isactrlsub {\isadigit{1}}{\isasymdiamond}\ k\ OR\ r\ {\isasymrestriction}\ {\isacharbraceleft}LENGTH{\isacharparenleft}key{\isacharparenright}{\isachardot}{\isachardot}{\isacharless}LENGTH{\isacharparenleft}word{\isadigit{3}}{\isadigit{2}}{\isacharparenright}\ {\isacharminus}\ LENGTH{\isacharparenleft}byte{\isacharparenright}{\isacharbraceright}{\isachardoublequoteclose}\smallskip}\\
Here the parameter $r$ represents some arbitrary memory state being overwritten by the representation of the capability. The binary representation of $r$ is truncated (by bit-wise conjunction with a mask) to fill the range of unused bits before combining it with the zero-padded representation. The value of unused memory $r$ is propagated across all representation functions in a composable way, so all low-level representations are formalized with plain single-valued functions. This approach not only allows for a simple transfer of all high-level properties to the low-level representation, but also avoids the need in explicit definitions of the corresponding inverse (decoding) functions. A single definition is enough to reuse the encoding functions (along with their injectivity proofs) for the specifications of operations that require decoding of representations:
\smallskip\\{
\noindent\isacommand{definition}\ \isamarkupfalse\isastyle%
{\isachardoublequoteopen}maybe{\isacharunderscore}inv\ f\ y\ {\isasymequiv}\ if\ y\ {\isasymin}\ range\ f\ then\ Some\ {\isacharparenleft}the{\isacharunderscore}inv\ f\ y{\isacharparenright}\ else\ None{\isachardoublequoteclose}
}\smallskip\\
Since we don't verify the actual implementation of the decoding functions, this implicit definition is sufficient and greatly simplifies proofs.

There is, though, one potential weakness in this approach in that it's still possible to accidentally lose some non-determinism when propagating the values of the unused memory by unintentionally identifying different values of the additional parameters. Each occurrence of the representation function should be provided its own separate instance of an additional parameter so that e.g. encoding of the whole kernel storage is supplied with the whole previous state of the store as an additional parameter rather than just a single additional default value. To systematically guarantee absence of such losses of non-determinism we prove additional lemmas of the form:
\smallskip\\{
  \isacommand{lemma}\isamarkupfalse\isastyle%
\ cap{\isacharunderscore}rep{\isacharunderscore}unused{\isacharcolon}\ {\isachardoublequoteopen}\mbox{{\isasymlfloor}c{\isasymrfloor}\ r\ {\isasymrestriction}\ unused\ {\isacharequal}} r{\isachardoublequoteclose},}\smallskip\\
where $unused$ is the set of unused bits and {\isasymrestriction} restricts the range of bits by zeroing out bits with indices not in the specified set. These lemmas, though, are proved very easily for all our representation function definitions.

\subsection{Disjointness of addresses}
Another problem arising from detailed low-level specifications of memory layout, such as the layout of the kernel storage, is the problem of reasoning about non-intersecting memory areas. While in the context of program verification there are such well-known approaches to reasoning about disjoint memory footprints as separation logic~\cite{reynolds_separation_2002} and dynamic frames~\cite{kassios_dynamic_2006}, in our context of formalizing the specification (rather than the implementation) of the system these approaches turned out to be both too abstract and too heavyweight. Too abstract since in separation logic the particular concrete layout of the memory footprints is left entirely abstract, while we needed to formalize the actual mapping of the data structures to the mostly fixed address ranges they should occupy. Too heavyweight since to represent the encoding of the whole kernel state with either separation logic or dynamic frames we would need to use some additional means to set up the embedding of the corresponding reasoning mechanism into plain HOL, while not having any real need in verifying code involving updates to the system state. In our approach we simply treated kernel addresses as semantic entities with some ascribed low-level representations (concrete values). Then following our general use of representation functions we defined the representation of addresses and its inverse. The inverse then can be directly used to specify the storage layout and prove the injectivity of the overall encoding with minimal effort. Here's an illustrative example:\\
{
\noindent\isacommand{typedef}\isamarkupfalse\isastyle%
\ offset\ {\isacharequal}\ {\isachardoublequoteopen}{\isacharbraceleft}\ n\ {\isacharcolon}{\isacharcolon}\ nat{\isachardot}\ n\ {\isacharless}\ {\isadigit{2}}\ {\isacharcircum}\ LENGTH{\isacharparenleft}byte{\isacharparenright}{\isacharbraceright}{\isachardoublequoteclose}\ \isakeyword{\small morphisms}\ off{\isacharunderscore}rep\ off\\
\noindent\isacommand{datatype}\isamarkupfalse%
\ address\ {\isacharequal}\ Nprocs \ {\isacharbar}\ Curr{\isacharunderscore}proc \ {\isacharbar}\ Proc{\isacharunderscore}heap\ offset\\
\noindent\isacommand{definition}\isamarkupfalse\isastyle%
\ {\isachardoublequoteopen}addr{\isacharunderscore}rep\ a\ {\isasymequiv}\ case\ a\ of\isanewline
\ \ \ \ Nprocs\ \ \ \ \ \ \ \ \ {\isasymRightarrow}\ {\isadigit{0}}x{\isadigit{0}}{\isadigit{0}}{\isadigit{0}}{\isadigit{0}}\isanewline
\ \ {\isacharbar}\ Proc{\isacharunderscore}heap\ offs\ {\isasymRightarrow}\ {\isadigit{0}}x{\isadigit{0}}{\isadigit{2}}{\isadigit{0}}{\isadigit{0}}\ OR\ of{\isacharunderscore}nat\ {\isacharparenleft}off{\isacharunderscore}rep\ offs{\isacharparenright}{\isachardoublequoteclose}\\
\noindent\isacommand{definition}\isamarkupfalse%
\ {\isachardoublequoteopen}addr{\isacharunderscore}inv\ {\isasymequiv}\ maybe{\isacharunderscore}inv\ addr{\isacharunderscore}rep{\isachardoublequoteclose}\\
\noindent\isacommand{definition}\isamarkupfalse\isastyle%
\ {\isachardoublequoteopen}encode\ {\isasymsigma}\ r\ a\ {\isasymequiv}\ case\ addr{\isacharunderscore}inv\ a\ of\isanewline
\phantom{mm} Some\ a{\isacharprime}\ \ \ \ \ \ \ \ \ \ \  {\isasymRightarrow}\ case\ a{\isacharprime}\ of\isanewline
\phantom{mmmm} Nprocs\ \ \ \ \ \ \ \ \ {\isasymRightarrow}\ of{\isacharunderscore}nat\ {\isacharparenleft}nprocs\ {\isasymsigma}{\isacharparenright}\ OR\ {\isacharparenleft}r\ a{\isacharparenright}\ {\isasymrestriction}\ {\isachardot}{\isachardot}{\isachardot} \isanewline
\phantom{mmm} {\isacharbar}\ Proc{\isacharunderscore}heap\ offs\ {\isasymRightarrow}\ encode{\isacharunderscore}heap\ {\isasymsigma}\ offs\ r\isanewline
\phantom{m} {\isacharbar}\ None\ \ \ \ \ \ \ \ \ \ \ \ \ \ \ \ \ {\isasymRightarrow}\ r\ a{\isachardoublequoteclose}
}\smallskip

Also note the filler of the unused memory $r$ being passed over in a top-down manner starting from the outermost representation function.

Now we move from the problems arising from the detailed low-level specification of our target system to some more general issues of formalization and formal proofs within the Isabelle/HOL framework that we encountered during verification.

\subsection{General Isabelle/HOL limitations}

\subsubsection{Bit-vector concatenation}

An example of a minor, though noticeable limitations of the simple Hindley-Milner type system employed within the Isabelle/HOL framework is its inability to express type-level sum (and other simple arithmetic operations), while still being able to express type-level numbers. For an illustration of the issue consider the following definition of bit-vector concatenation function from the HOL-Word library that comprises an extensive Isabelle/HOL formalization of fixed-size bit-vectors, corresponding operations and their various properties:

\smallskip{
\noindent\isacommand{definition}\isamarkupfalse\isastyle%
\ word{\isacharunderscore}cat\ {\isacharcolon}{\isacharcolon}\ {\isachardoublequoteopen}{\isacharprime}a{\isacharcolon}{\isacharcolon}len{\isadigit{0}}\ word\ {\isasymRightarrow}\ {\isacharprime}b{\isacharcolon}{\isacharcolon}len{\isadigit{0}}\ word\ {\isasymRightarrow}\ {\isacharprime}c{\isacharcolon}{\isacharcolon}len{\isadigit{0}}\ word{\isachardoublequoteclose}\ ...
}\smallskip

The annotation of the form \textit{{\isacharprime}a{\isacharcolon}{\isacharcolon}len{\isadigit{0}}} constrains the type parameter \textit{'a} to belong to the \textit{len0} type class, which has the corresponding associated operation \textit{LENGTH('a)} returning a natural number. Thus we essentially gen type-level numbers that can be injected into terms as natural numbers with the use of the \textit{LENGTH} operation. However, as we can see in the definition of \textit{word\_cat}, the result of this function has the type \textit{'c::len0} that is generally unrelated to the parameter types \textit{'a} and \textit{'b}. This has two basically unavoidable, but undesirable consequences:
\begin{itemize}
    \item Since there is no way of further constraining the resulting parameter type \textit{'c::len0}, the function \textit{word\_cat} is forced to be partial. Generally, there is nothing particularly special about handling of partial functions within the Isabelle/HOL framework, but their presence has at least one undesirable consequence for formalization of system interface specifications, which we discuss further in this section.
    \item Since the resulting type parameter \textit{'c::len0} cannot be automatically inferred from the arguments of \textit{word\_cat}, if has to be explicitly specified. Normally, this doesn't lead to a significant type annotation burden since the parameter can be propagated by type inference from some term with a known type. But in case of consecutive (nested or chained) \textit{word\_cat} applications, the inner type parameters become essentially inaccessible for further type propagation or inference and have to be specified explicitly e.g.\\
    \noindent{
    \noindent\isacommand{definition}\isamarkupfalse\isastyle%
\ {\isachardoublequoteopen}entry{\isacharunderscore}proc{\isacharunderscore}addr\ {\isasymequiv}\ word{\isacharunderscore}cat\ \ \isanewline
\ \ \ \ \ {\isacharparenleft}word{\isacharunderscore}cat
\ {\isacharparenleft}word{\isacharunderscore}cat\ {\isacharparenleft}k{\isacharunderscore}prefix\ {\isacharcolon}{\isacharcolon}\ {\isadigit{3}}{\isadigit{2}}\ word{\isacharparenright}\ {\isacharparenleft}{\isadigit{0}}x{\isadigit{0}}{\isadigit{4}}\ {\isacharcolon}{\isacharcolon}\ byte{\isacharparenright}\ {\isacharcolon}{\isacharcolon}\ {\isadigit{4}}{\isadigit{0}}\ word{\isacharparenright}\isanewline
\ \ \ \ \ \ \ {\isacharparenleft}{\isadigit{0}}\ {\isacharcolon}{\isacharcolon}\ {\isadigit{1}}{\isadigit{9}}{\isadigit{2}}\ word{\isacharparenright}\ {\isacharcolon}{\isacharcolon}\ {\isadigit{2}}{\isadigit{3}}{\isadigit{2}}\ word{\isacharparenright}
\ {\isacharparenleft}{\isadigit{0}}x{\isadigit{0}}{\isadigit{0}}{\isadigit{0}}{\isadigit{0}}{\isadigit{0}}{\isadigit{0}}\ {\isacharcolon}{\isacharcolon}\ {\isadigit{2}}{\isadigit{4}}\ word{\isacharparenright}\ {\isacharcolon}{\isacharcolon}\ {\isadigit{2}}{\isadigit{5}}{\isadigit{6}}\ word{\isachardoublequoteclose}}
\end{itemize}

This can be slightly mitigated by introducing some ad-hoc monomorphic notation for hexadecimal numbers (e.g. syntactically reconstructing the type annotation from the length of the input hexadecimal representation), but this approach still quickly becomes unwieldy in practice, especially in the context of the great available variety of Ethereum bit-vector types with various lengths.

First we propose a relatively simple remedy for the second problem. We actually used our own definition of a concatenation function with a fixed result type (the largest needed length of 256 bits) and parameter types of arbitrary length that is ignored. Instead we provided the necessary length of the second argument as an additional explicit parameter. Thus the whole issue of dealing with lengths was shifted from the type to the term level eliminating the need in any type-level representations altogether. This resulted in more approachable definitions e.g.

\smallskip{
\noindent\isacommand{definition}\isamarkupfalse%
\ {\isachardoublequoteopen}entry{\isacharunderscore}proc{\isacharunderscore}addr\ {\isasymequiv}\isanewline
\ \ {\isacharparenleft}k{\isacharunderscore}prefix\ {\isacharcolon}{\isacharcolon}\ {\isadigit{3}}{\isadigit{2}}\ word{\isacharparenright}\ {\isasymJoin}\isactrlbsub {\isadigit{2}}{\isadigit{2}}{\isadigit{4}}\isactrlesub \ {\isadigit{0}}x{\isadigit{0}}{\isadigit{4}}\ {\isasymJoin}\isactrlbsub {\isadigit{2}}{\isadigit{1}}{\isadigit{6}}\isactrlesub \ {\isacharparenleft}{\isadigit{0}}\ {\isacharcolon}{\isacharcolon}\ {\isadigit{1}}{\isadigit{9}}{\isadigit{2}}\ word{\isacharparenright}\ {\isasymJoin}\isactrlbsub {\isadigit{2}}{\isadigit{4}}\isactrlesub \ {\isadigit{0}}x{\isadigit{0}}{\isadigit{0}}{\isadigit{0}}{\isadigit{0}}{\isadigit{0}}{\isadigit{0}}{\isachardoublequoteclose}
}\smallskip\\
Here $\cdot \isasymJoin_\cdot \cdot$ denotes our concatenation function. In our opinion in the lack of dependent types or other expressive capabilities of the type system the use of logical (term-level) constraints may be often preferable to some limited meta-logical (e.g. type-level) extensions such as the use of type classes. Now we move to the second problem.

\subsubsection{Partiality}

The presence of partial functions in the specification of an interface of the system has a subtle undesirable property~--- unpredictability stemming from the undefined results returned by the partial functions. Consider the following very typical and general preservation lemma:

\smallskip{
\noindent\isacommand{lemma}\isamarkupfalse\isastyle%
\ preservation{\isacharcolon}\ {\isachardoublequoteopen}I\ s\ {\isasymLongrightarrow}\ I\ {\isacharparenleft}op\ s\ a{\isacharparenright}{\isachardoublequoteclose}}\smallskip

Here \textit{I} is an invariant of the system and \textit{op} is an operation on the system with an argument \textit{a}. Let's imagine an example instance of this kind of lemma: Let $s$ be a natural number, \textit{I s} be the predicate $s > 0$ and \textit{op} correspond to the operation $ s \gets s + s \mathop{\mathrm{div}} a$. Looking at the general statement of the lemma, a rather natural interpretation of such a preservation property would be that any application of the operation \textit{op} to the system is ``safe'' as it preserves its invariant. However in our particular example it's obvious that even though the application of \textit{op} with $a = 0$ provably preserves the invariant, it actually has entirely unpredictable consequences for the system. So specifications of operations on the system involving partial functions may considerably mislead the reader of the specification while remaining perfectly correct form the purely logical perspective. If the formal specification is to serve as a formal documentation on the system this fact may significantly undermine the value of applying the formal methodology for that purpose. Fortunately, there are various ways to strengthen the specification to exclude such unintuitive definitions. For our specification we additionally proved the following injectivity-like lemmas for every operation:

\smallskip{
\noindent\isacommand{lemma}\isamarkupfalse\isastyle%
\ injectivity{\isacharunderscore}like{\isacharcolon}\ {\isachardoublequoteopen}op\ s\ a\ {\isacharequal}\ op\ s\ b\ {\isasymLongrightarrow}\ a\ {\isachartilde}\ b{\isachardoublequoteclose}}\smallskip\\
Here $\sim$ denotes some notion of equivalence for arguments of the operation in a sense that equivalent arguments produce equivalent results. In case the operation $op$ actually involves some non-determinism, the formulation of the lemma should be adjusted accordingly, thus making this non-determinism explicitly exposed for the reader. The proof of such a lemma is enough to exclude any hidden non-determinism, since for any non-trivial equivalence relation $\sim$ ($\exists a'.~a' \not\sim a$) if the \textit{op} has non-deterministic result on $a$, $op~a$ may be arbitrarily chosen to be equal to $op~a'$ and the relation $a \sim a'$ then cannot be established.

\subsubsection{Dependent products}
Another limitation arising from the lack of dependent types or other expressive type system features is inability to directly express dependent products i.e. types of the form $\prod_{x :: 'a}{f(x)}$, where $f$ is a type-level function on the value $x$ of some type $'a$. A typical example of a situation, where this seems very natural is a list of pairs of the form ``\textit{capability\_type} $\times$ \textit{capability\_representation}'' (e.g. if the value of the first member is ``\textit{Write}'', than the type of the second member should be ``\textit{write\_capability}''). Such types cannot be directly expressed within the Isabelle/HOL framework. A typical workaround is to use injection into some universal type with additional well-formedness predicates stated as preconditions to the operations or as type invariants. Here we were able to directly reuse our representation functions for injecting different types of capabilities into the same universal bit-level representation.

Finally it's important to note an essential benefit of a logical framework with a very limited type system, which is its amenability to automation using existing readily available tools such as saturation-based provers (E-prover, Vampire, Metis) and SMT solvers (Z3, CVC4). In our experience their use within the Isabelle framework lead to great advantages ultimately outweighing all the limitations mentioned above. Overall, the formalized specification with proofs took about 4500 lines of Isabelle based on 25 pages of the original textual description.

\section{Related Work}
There are many examples of using formal methods for developing specifications of various systems.
Isabelle/HOL was used to prove functional correctness of the seL4 operating system microkernel~\cite{klein_comprehensive_2014}, providing a proof chain from the high-level abstract specification of the kernel, down to the executable machine code.
The B-method was applied to create formal models of various safety-critical railway systems~\cite{lecomte_formal_2007}.
A dedicated specification language for defining the high-level abstract models was introduced in~\cite{chaudhuri_practical_2016}.

On the other hand, verification of smart contracts is almost exclusively concentrated on the contract implementation, omitting the separate formalisation of their specification.
It is a valid approach if the specification is simple enough, which is not the case for the Cap9 framework.

There are several examples of formalisation of the Ethereum virtual machine: using the K framework~\cite{hildenbrandt_kevm:_2018}, the Lem language~\cite{hirai_defining_2017}, F*~\cite{grishchenko_semantic_2018}, and Isabelle~\cite{amani_evm_semantics}, which can serve as a basis for formal verification of the contract code.
Why3 platform for deductive program verification was recently applied for writing and verifying smart contracts~\cite{nehai_deductive_2019}.

\section{Conclusion and Future Work}
We have developed a formal specification\footnote{The specification is publicly available at \url{https://github.com/Daohub-io/cap9-spec}} of the Cap9 framework using the Isabelle/HOL theorem prover and proved its internal consistency.
To create it we have employed a refinement approach based on representation functions, which allowed us to efficiently use powerful automated reasoning tools provided with Isabelle.
We have found Isabelle/HOL to be suitable for developing specifications of smart contracts, although some minor issues were identified and outlined.

The next step is formal verification of the Ewasm implementation of the Cap9 framework for its compliance with the Isabelle/HOL specification, which may require developing some additional tools.
Other possible direction is to develop and verify a higher level permission system that is based on the Cap9 primitives.

\bibliographystyle{splncs04}
\bibliography{bibliography}

\end{document}